\documentclass[11pt,twoside]{article}

\usepackage{asp2004}
\usepackage{epsf}
\usepackage{graphicx}
\usepackage{psfig}
\usepackage{lscape}

\begin{document}
\title{The Secondary Standards programme for OmegaCAM at the VST}   
\author{Gijs Verdoes Kleijn, Ronald Vermeij \& Edwin Valentijn}   
\affil{OmegaCEN, Kapteyn Astronomical Institute, Groningen, The Netherlands}    
\author{Koen Kuijken}   
\affil{Leiden Observatory, Leiden, The Netherlands}    

\begin{abstract} 
The OmegaCAM wide-field imager will start operations at the
ESO VLT Survey Telescope at Paranal, Chile, in 2007. The photometric calibration of OmegaCAM data depends on standard star measurements that cover the complete
1 square degree FoV of OmegeaCAM. A catalog fullfilling this requirement for eight Landolt equatorial fields, denoted the OmegaCAM Secondary Standards Catalog, will be
constructed from OmegaCAM observations during the first year of operations. Here we present the 'Preliminary Catalog' which will be used to bootstrap the construction of the OmegaCAM Secondary Standards Catalog. Thus the Preliminary Catalog will be used to assess the performance of OmegaCAM+VST early-on. The catalog is based on WFC data from the INT at La Palma.

\end{abstract}


\section{Introduction}   
The number of wide field optical imagers with a large FoV ($\geq 1$ square degree) available at astronomical telescopes is growing.
These instruments perform surveys which can cover thousands of square degrees of sky aimed at a wide range of Galactic and extragalactic science goals. The surveys are made public often promptly allowing astronomers from around the globe to use the data for their own specific scientific needs. The imagers contain many CCDs and often elaborate components to correct for optical distorsions over the FoV. Accurate and robust photometric calibration over the complete area of such surveys is required to successfullly use the data for the many different (and sometimes unforeseen) science projects.
In this article we discuss the photometric calibration of a new wide-field imager: OmegaCAM. 

First we discus the main properties of the instrument and its telescope.
OmegaCAM has been constructed by a consortium consisting
of Leiden Observatory, Kapteyn Astronomical Institute, the University Observatories of Munich, G\"{o}ttingen and Bonn,
the Padova and Capodimonte Astronomical Observatories and
the European Southern Observatory (ESO). It has an array of 32 thinned, low-noise (5e$^{-}$) 2x4k E2V CCDs which cover an area of 1x1 deg$^{2}$ at 0.21 arcsec/pixel (Kuijken et al.\ 2004).
The key filters used will be the Sloan $u$, $g$, $r$, $i$ and $z$
filters (see Figure~\ref{f:filters} for measured throughputs). Besides this, the list of filters includes a filter segmented into a $u$, $g$, $r$, and $i$ quadrant, various segmented narrow-band filters and Johnson $B$ and $V$ filters. The OmegaCAM wide-field imager will be the sole instrument on the ESO VLT Survey Telescope (VST, Capaccioli et al.\ 2005) and will be mounted at the Cassegrain focus.
The VST is a 2.6m telescope being built by ESO and the Capodimonte Astronomical Observatory. 
Its modified Ritchey-
Cr\'{e}tien set-up is specifically designed for wide-field imaging, and
has been optimized for excellent image quality in natural seeing.
Three large ESO Public surveys (see http://www.eso.org/observing/webone.html) and many guaranteed-time surveys will be carried out covering thousands of square degrees in total. The surveys address Galactic science (e.g., ultra-compact binary systems, transiting planets) and extragalactic science (e.g., gravitational weak lensing, galaxy evolution in superclusters). Telescope and instrument are expected to start operations at Paranal, Chile in 2007. OmegaCAM is expected to have a data rate which is equivalent to the rate for observing the Southern hemisphere in about three years through one filter. This will result in over hundred Terabyte of raw plus processed data.      

\begin{figure}
\caption{The measured filter throughputs for the $g'$ (in green), $r'$ (in red), $i'$ (in yellow) and $z'$ (in black) filters for OmegaCAM. The filter+CCD throughputs are also indicated.}
\label{f:filters}
\end{figure}

\section{The photometric calibration of OmegaCAM}

A prime concept of the OmegaCAM calibration plan is to derive the photometric calibration for each of the 32 chips completely independently. This will improve the photometric accuracy compared to a method which averages over (a subset of) the chips. Another important concept of the OmegaCAM photometric calibration is to continuously maintain the
photometric scale in the Sloan bands, even when the science programme does not require the
usage of these passbands during a particular night or period. This continuity will ease maintaining an accurate photometric system. We plan to use a scalable standard extinction curve which facilitates (i) recognizing deviating data, (ii) processing of incomplete data and (iii) performing quality checks. OmegaCAM will monitor the atmosphere, in partciular its extinction, via observations of a fixed near-polar field about three times per night. The near-polar field is observable throughout the year. Nightly observations of standard star fields then allow accurate photometric calibration. Further details on the photometric calibration plan for OmegaCAM can be found in the calibration plan manual available at http://www.astro.rug.nl/~omegacam/documents.

The result of this pipeline is that, during automated processing, the photometric calibration of the broad-band filters of OmegaCAM achieves the required accuracy of 5\% or better on the photometric scale in `instrumental magnitudes' as assigned
to the units of a resultant output image of the image pipeline. In dedicated processing the accuracy is expected to be 2\% or better. The accuracy of the colour
transformation terms of instrumental to standard systems will be better than 10\%. Another relevant number is the accuracy of flats. Laboratory tests have shown that the internal calibration unit for dome flats has a stability which is better than 1\% over at least a period of 4 days. The unit lamp is a type no.\ 14612 Philips Brilliant lamp produced in Poland. Tests have shown that, with the help of this high stability, domeflats can be used even to determine accurately ($<0.3\%$) chip-to-chip variations in the gain.

A photometric calibration for each chip independently requires a spatially dense coverage of standard stars which covers the complete
1 square degree FoV (not the least to be efficient in terms of observing time). Catalogs of photometric standard stars that meet this requirement do not exist currently. We will obtain eight of such 1 square degree fields of secondary standards, each centered on a Landolt equatorial field, during the first year of OmegaCAM operations (see Table~\ref{t:standard_fields}). The result will be the OmegaCAM Secondary Standards catalog. The chosen approach is to observe each of the eight Landolt fields centered on
each individual OmegaCAM CCD. These Landolt equatorial fields already contain standards measured by Landolt (1992) and by Stetson (2000) in the $UBVR_{\rm c}I_{\rm c}$ Johnson-Cousins system and measurements from Data Release 5 (DR5, Adelman-McCarthy 2007 in the Sloan $u,g,r,i,z$ system. The observations will also be used to detect and assess effects that influence the photometric accuracy as a function of position on the chip, e.g., illumination variations. This poses a bootstrap problem: determining
accurate zeropoints of the secondary standards requires that an illumination correction is
already known. To address this bootstrap problem we need to obtain a quick and efficient initial determination of the illumination correction for OmegaCAM based on a preliminary catalog of standard stars. We have constructed this Preliminary Catalog from observations of 1 square degree on the equatorial Landolt fields using the WFC at the INT on La Palma.

\section{The Preliminary Catalog}

\subsection{Observations and data reduction}

The WideField-Camera (WFC) on the 2.5m Isaac Newton Telescope at La
Palma is a mosaic camera consisting of four CCDs covering an area of
about $0.5^o \times 0.5^o$ with a pixel scale of
$0.333\arcsec$. The runs were performed in June 2002
and February 2003. All
Landolt fields listed in Table~\ref{t:standard_fields} except SA95
were observed.  Every Landolt field was observed in 5 partially overlapping pointings covering $\sim 1.1^o \times 1.1^o$ on the sky. To cover an
adequate range of magnitudes both a short ($\sim
10s$) and a long exposure ($\sim 300s$) were made for each pointing. All the
observations were done in the $u$, $g$, $r$, $i$ and $z$ bands.

\begin{table*}
\caption{The location (J2000) of the eight standard star fields. All fields in the list cover an area of 1.1$\times$1.1 degree
centered on the coordinates given. Also listed are the number of stars obtained for the Preliminary Catalog (PC) from the WFC INT observations, the number of Landolt and Stetson standards and the number of Sloan DR5 stars for each field.}
\label{t:standard_fields}\begin{center}
\begin{tabular}{lccrrrr}
\hline
Field           & alpha & delta & \#PC & \#Landolt & \#Stetson & \#SDSS DR5 \\
                & (deg)    & (deg)  & & & & \\
 \hline\\
SA\,92         &   13.946 &  $+$0.949 & 6475  & 41 &   210 &  882   \\
SA\,95         &   58.500 &  $+$0.000 & 0     & 45 &   417 &  1154  \\
SA\,98          &  103.021 &  $-$0.328 & 23840 & 46 &   1116&  0     \\
SA\,101        &  149.112 &  $-$0.386 & 5591  & 35 &   117 &  1842  \\
SA\,104        &  190.592 &  $-$0.553 & 5701  & 34 &   76  &  1964  \\
SA\,107        &  234.896 &  $-$0.252 & 12006 & 28 &   728 &  4143  \\
SA\,110         &  280.679 &  $+$0.348 & 38562 & 39 &   589 &  0     \\
SA\,113        &  325.533 &  $+$0.493 & 13947 & 42 &   477 &  4044  \\
\hline
\end{tabular}\end{center}
\end{table*}

The WFC data has been reduced using the Astro-Wise Environment (Valentijn \& Kuijken 2001, Valentijn \& Verdoes Kleijn 2006). 
The scientific exploitation of many of the OmegaCAM surveys will be carried out using the Astro-Wise Environment . The Astro-Wise Environment provides a novel way to deal intelligently with the large amount of data coming from optical wide field imagers in general and OmegaCAM in particular. It is a fully scalable 'science information system' which unifies the archiving, the processing and advanced analysis tools. The hardware and software that make up the environment are internationally federated facilitating collaborators in many places to share,
validate and combine processed data and to pool hardware resources. All information about raw and reduced frames and the
derived results from analysis on them are stored in catalogues in the federated database which constitutes the heart of the system.
The Astro-Wise environment has been developed in the last five years by a partnership between OmegaCEN-NOVA/
Kapteyn Institute (coordinator),
Osservatorio Astronomico di
Capodimonte, Terapix at IAP (France),
ESO, Universit\"{a}ts-Sternwarte \&
Max-Planck Institut f\"{u}r
Extraterrestrische Physik
(Munich, Germany).

The WFC data reduction consisted of the
usual steps of de-biasing (no overscan was used), flat-fielding and astrometric calibration. The masterflat was constructed
from domeflats and/or twilightflats. Defringing has been applied on the $z$ band data.
No illumination variation has been detected in accordance with other users of WFC (M. Irwin private communication).
One of the WFC chips (A5506-4) turned out to have a measurable non-linear countrate in agreement with the non-linearity report by the CASU INT Wide Field Survey (McMahon et al.\ 2001). The non-linearity turned out to be too weak to affect the overall photometric calibration and hence no correction was made.
Finally, the astrometry was determined per individual frame using the USNO catalog.

The aperture flux measurements and its errors were obtained in an automated fashion using SExtractor with a weight frame. The stellar fluxes were measured through a circular aperture with a diameter of 30 pixels ($\sim 10''$). Stars not suited for accurate photometry were removed from the list (e.g., saturated stars, stars with overlapping apertures, stars on the edge of chips). After this selection many thousands of stars remained. The number of stars in the Preliminary Catalog and the various literature sources are given in Table~\ref{t:standard_fields}. 

\subsection{Photometric calibration}

The goal is to put the WFC data for the Preliminary Catalog on the Sloan photometric system. 
Ideally one would like to tie the standard in the Preliminary Catalog to the fundamental standards that define the Sloan photometric system. Unfortunately, the sparseness of the Sloan standards in the Landolt equatorial fields and in the Southern Hemisphere did not allow us to do this within reasonable observing time. Moreover the brightness of these standards poses technical problems as well. 

Therefore we followed the strategy of observing Landolt fields which contain both Landolt standards in the Johnson-Cousins photometric system and stars in the Sloan Digital Sky Survey. We photometrically calibrate the WFC data twice: once with the Landolts stars as calibrators and once with the DR5 stars as calibrators. The difference in the results puts constraints on the systematic and random errors in the photometric zeropoints as will be discussed below. It will turn our that we cannot prefer clearly one calibration over the other at this point. In the end we decided, for mostly practical reasons, to use the calibration based on DR5 for the Preliminary Catalog.

For the Landolt calibrator set we use the transformations determined by Jester et al.\ (2005) to convert the $UBVRI$ magnitudes to the Sloan $u,g,r,i,z$. This transformation is based on a comparison of magnitudes for standards that are used in the definition of both photometric systems. In this way, the photometric calibration of the WFC data with Landolt stars is indirectly tied to the Sloan standards.
Using the stars from SDSS DR5 as calibrators offers many more stars (see Table~\ref{t:standard_fields}) but far less accurate individual calibrator magnitudes. However, the Sloan photometry of DR5 is much more directly tied to the standards that define the Sloan photometric system (Tucker et al.\ 2006).  

To convert the WFC natural system to the Sloan $u,g,r,i,z$ photometric system we solve the following equation for each filter:
\begin{equation}
M_i^{\rm CAL} -m_i^{\rm WFC}=Z - k X +CT \times (M_i^{\rm CAL}-M_j^{\rm CAL}),
\end{equation}
where $M_i^{\rm CAL}$ is the Sloan magnitude of the calibrator (either from SDSS DR5 or the Landolt catalog) in filter $i$, $m_i^{\rm WFC}$ is our instrumental magnitude, $Z$ is the zeropoint, $k$ the atmospheric extinction coefficient, $X$ the airmass, $CT$ the color coefficient and $M_i^{\rm CAL}-M_j^{\rm CAL}$ the color defined by filter $i$ and $j$. 
The following fixed values were used for the atmospheric extinction coefficients:
$k_{u,g,r,i,z}=[0.47,0.19,0.09,0.05,0.05]$. These values are based on the values derived by the CASU INT Wide Field Survey (McMahon et al.\ 2001). They agree within the errors for the determination on our observing nights as done by the Carlsberg Meridian Telescope on La Palma (only $r$ band measurements).

Color terms for all bands have been determined for our data. The analysis
was based mainly on catalogs derived from the data of all the five pointings in the fields SA101 and
SA107 (2003 run) and SA 113 (2002 run). 
The color terms were all found to be small or even non-existent. This is not surprising,
because the Sloan filters at the INT/WFC closely match the original ones. The color terms for
$g$ and $i$ are $0.14 \pm 0.01$ and $0.07 \pm 0.01$, respectively, using ($g-r$) and ($r-i$) as the colors. No
evidence for any color term was found for the $u$, $r$ and $z$ bands. The zeropoint 
$Z$ was determined by taking the average of the typically 10 zeropoints from individual observations of the fields (i.e., 5 pointings $\times$ (1 short+1 long exposure)). We removed outliers among them.  
Finally, a star was admitted to the standard star catalog provided it had a $r$ magnitude and magnitudes for at least two more Sloan bands. The $r$ magnitude distribution of the resulting catalog is shown in Figure~\ref{f:Aw2sRmags}. 

\begin{figure}
\centering
\caption{The distribution of Sloan $r$ magnitudes for the WFC Secondary Standards.}
\label{f:Aw2sRmags}
\end{figure}

\subsection{Random errors}
\label{s:randomerrors}

Random errors in the photometric calibration arise partly from random errors in the calibrator sets.
The scatter in the zeropoints of different Sloan fields, which are 0.03 square degrees, for the photometric calibration of the DR5 itself is estimated to have a rms width of $u\sim 0.03$mag, $\sim 0.01$mag in $r,g,i$ and $z \sim 0.02$mag (Ivezi{\'c} et al.\ 2004). For the photometric calibration based on the Landolt catalog a main source of random errors is expected to be the usage of transformation equations. The equations from Jester et al.\ (2005) were derived from stars which have $R_{\rm c}-I_{\rm c}<1.15$. They report a rms scatter in the conversion of individual stars of 0.06mag in $u-g$, 0.04mag in $g-r$, 0.03mag in $r-i$, 0.03mag in $r-z$, 0.02mag in $g$ and 0.03mag in $r$. These transformation equations made also use of the very small transformation between the Sloan $u',g',r',i',z'$ photometric system (on which the Sloan fundamental standards were observed) and the $u,g,r,i,z$ Sloan photometric system (on which the DR5 stars were observed; see Tucker et al.\ 2006). The rms scatter in this transformation seems to be $\sim 0.015$mag typically. It is difficult to determine how these errors for individual stars fields / small patches of sky propagate into zeropoint errors for the WFC data. In fact, a more direct estimate of the overall random error in the zeropoint of the instrumental magnitudes per CCD ($\sim 0.06$ square degrees) can be obtained from the scatter in the individual zeropoints determined per chip plus filter combination over the different fields throughout the photometric part of the night. The measured $1\sigma$ scatter in these is $u \sim 0.020$mag, $g \sim 0.015$mag, $r \sim 0.015$mag, $i \sim 0.015$mag and $z \sim 0.020$mag. 

\subsection{Systematic errors}

We estimated the systematic error of our photometric calibration by comparing the Sloan magnitudes for the $\sim 30$ stars in the eight Landolt equatorial fields which appear both in SDSS DR5 and in the Landolt catalog. Table~\ref{t:SloanVsLandolt} shows that the residuals of the magnitudes indicate systematic differences $<0.05$mag except for the $u$ band.
\begin{table*}
\caption{The residuals in the Sloan photometric system for 32 stars in the eight Landolt fields which appear both in SDSS DR5 and in the Landolt catalogue. The Landolt Johnson-Cousins magnitudes were converted to the Sloan photometric system using the transformations given by Jester et al.\ (2005). The residual is defined as the magnitude from SDSS DR5 minus the one from Landolt.}
\label{t:SloanVsLandolt}
\begin{center}
\begin{tabular}{c r c}        
residual & median & uncertainty \\
$\Delta u$ & 0.07  & 0.02 \\ 
$\Delta g$ & 0.02  & 0.01 \\
$\Delta r$ & -0.03 & 0.01 \\
$\Delta i$ & -0.04 & 0.02 \\
$\Delta z$ &  0.00 & 0.04  \\
\end{tabular}     
\end{center}
\end{table*}

Another estimate of the systematic error in the photometric calibration is to compare directly the zeropoints as determined using the Landolt catalog and DR5 catalog respectively. This is effectively an independent estimate from the previous one as the $\sim 30$ sources which appear in both catalogs form a very small subset of both catalogs. 
The resulting residuals in zeropoint, defined as the zeropoint using DR5 stars minus the one based on the Landolt catalog have the following mean values: $0.05$mag in $u$, $0.00$mag in $g$, $-0.02$mag in $r$, $ -0.05$mag in $i$ and $0.00$mag in $z$. Comparing these results to those in (Table~\ref{t:SloanVsLandolt} we see that both methods yield very similar results. 
 
Given that the estimates of the systematic error with and without using the WFC observations are very similar it seems likely that the systematic error is due to inconsistencies between the DR5 photometric calibration and the calibration via the Landolt catalog plus the transformation equations from Jester et al.\ (2005). 
Only a handful of stars in the Landolt catalog are present typically on a CCD. However it seems very unlikely that the systematic error, present in eight Landolt fields, has its origin in small number statistics given the small size of the random errors (see Section~\ref{s:randomerrors}) relative to the systematic error estimates. It is more likely that for example the usage of a single set of transformation equations might contribute to the systematic error.    
In the end, it turns out that we cannot prefer clearly one calibration over the other. We decide, for mostly practical reasons, to use the calibration based on DR5 for the Preliminary Catalog for the early verification of OmegaCAM+VST. Which calibrator set to use for the OmegaCAM Secondary Standards Catalog remains to be determined.
 
\section{Analysis of the Preliminary Catalog}

We inspected the residuals in the magnitudes obtained from the WFC data and the SDSS DR5 catalog for overlapping stars in the five Landolt fields which contain measurements of both sources. Small constant offsets ($\sim 0.03$) between the magnitudes were noted in some filters in some fields which were not used for the determination of the zeropoints. From inspection of the stellar locus on color-color plots we conclude that the SDSS DR5 most likely has consistent photometric calibration over all Landolt fields. Thus we ascribe the small constant offsets due to initially unnoticed atmospheric variations during our WFC observations and correct the magnitudes from the WFC data to bring them into accordance with the DR5 data. No systematic trends as function of e.g.\ magnitude or position have been detected once these constants have been applied. Figure~\ref{f:Aw2sVsSDSS} shows the residuals as a function of magnitude. For $u$ the residuals between the WFC and DR5 measurements are as expected from the measurement errors estimated for both magnitudes. For the other bands the scatter in the residuals is typically 2 times that expected on the basis of our measurement errors. Given the simplicity of our aperture photometry and error computation for the WFC data we suspect that this is due to underestimation of the error in our data. Figure~\ref{f:colorVScolor} shows the color distributions for the DR5 stars and the Preliminary Catalog stars. These figures illustrate the larger random error in the Preliminary Catalog stars. However, overall no systematic offsets are present. 

\begin{figure}
\caption{The residuals between our WFC Secondary Standards and the SDSS DR5 (psfMags) as a function of magnitude. The statistics for the whole sample of overlapping stars is listed.}
\label{f:Aw2sVsSDSS}
\end{figure}

\begin{figure}
\caption{The various color-color plots for the stars in the Preliminary Catalog (grey symbols) and in the SDSS DR5 (black symbols) for fields that contain stars from both catalogs. There are no systematic offsets but the scatter in the Preliminary Catalog data is larger.}
\label{f:colorVScolor}
\end{figure}

\section{Conclusions}

The main goal of the Preliminary Catalog is to verify the performance of OmegaCAM+VST and in this way bootstrap the construction of the catalog of Secondary Standard Stars from OmegaCAM observations in the first year of operations. Figures~\ref{f:Aw2sVsSDSS} and \ref{f:colorVScolor} indicate that we will be able to use the WFC data set to statistically detect and quantify photometric effects as e.g., a function of location on the detectors for OmegaCAM+VST. The analysis discussed here shows that the random error in the zeropoint of instrumental magnitudes using the Preliminary Catalog is $<0.05$mag. The systematic errors involved in putting the WFC data on the Sloan photometric system are $<0.1$mag. The scatter in individual magnitudes from the WFC observations is too large to have the Preliminary Catalog stars qualify as 'secondary standards' in the usual sense. The observations of the Landolt fields with OmegaCAM should be able to achieve the level of accuracy required to establish the catalog of OmegaCAM Secondary Standard Stars. 

\acknowledgements 

Part of the analysis and of the figures presented here was made using the TOPCAT software developed by Mark Taylor (www.starlink.ac.uk/topcat).


\end{document}